\documentclass[twocolumn,letterpaper,nofootinbib,amsmath]{revtex4}

\pdfoutput=1
\usepackage{epsfig}
\usepackage{hyperref}

\newcommand{\lessthansimilarto}{\lower3pt\hbox{$\buildrel{<}\over{\sim}$}}
\newcommand{\greaterthansimilarto}{\lower3pt\hbox{$\buildrel{>}\over{\sim}$}}

\newcommand{\RR}{\hbox{$I$\kern-3.8pt $R$}}

\begin{document}

\title{Puncture Evolution of Schwarzschild Black Holes}

\author{J.~David Brown}
\affiliation{Department of Physics, North Carolina State University,
Raleigh, NC 27695 USA}

\begin{abstract}
The moving puncture method is analyzed for a single, non-spinning black 
hole. It is shown that the puncture region is not resolved by current 
numerical codes. As a result, the 
geometry near the puncture 
appears to evolve to an infinitely long 
cylinder
of finite areal radius. The puncture 
itself actually remains at spacelike infinity throughout the evolution. 
In the limit of infinite resolution the data never become stationary. 
However, at any reasonable finite resolution the grid points closest 
to the puncture are rapidly drawn into the black hole interior by the 
$\Gamma$--driver shift condition. The data can then evolve to a 
stationary state. These results suggest that the moving puncture 
technique should be viewed as a type of ``natural excision". 
\end{abstract}
 
\maketitle
With the moving punctures technique, vacuum black holes are modeled on a numerical 
grid with $\RR^3$ topology. The interior of each black hole contains an asymptotic region that is 
represented as a point or ``puncture''  where the gravitational field diverges. 
To be precise, the initial data are chosen to be conformally flat, $g_{ab} = \psi^4 {\tilde g}_{ab}$, 
where ${\tilde g}_{ab}$ is the flat metric in Cartesian coordinates. The conformal factor $\psi$ behaves 
like $1/r$ at each puncture where $r$ is the coordinate distance from the puncture point. The numerical 
grid is constructed in such a way that the punctures do not coincide with any grid point. The initial data 
are evolved with the BSSN formulation of the evolution equations \cite{Shibata:1995we,Baumgarte:1998te}
along with  ``1+log'' slicing, 
\begin{equation}\label{onepluslog}
  \frac{\partial\alpha}{\partial t}  =  \beta^a\partial_a\alpha - 2\alpha K \ ,
\end{equation}
and the ${\Gamma}$--driver shift condition, 
\begin{subequations}\label{gammadriver}
\begin{eqnarray}
  \frac{\partial\beta^a}{\partial t} & = & \frac{3}{4} B^a \ ,\\
  \frac{\partial B^a}{\partial t} & = & \frac{\partial {\tilde\Gamma}^a}{\partial t} - \eta B^a \ .
\end{eqnarray}
\end{subequations}
Here, $\alpha$ is the lapse function, $\beta^a$ is the shift vector, and $K$ is the trace of the extrinsic 
curvature. The 
conformal connection functions are defined by ${\tilde\Gamma}^a \equiv - \partial_b {\tilde g}^{ab}$ 
with $\det({\tilde g}_{ab}) = 1$. Variants of these gauge conditions  are obtained by  
dropping the advection term $\beta^a\partial_a\alpha$ in Eq.~(\ref{onepluslog}) or replacing one 
or more of the time derivatives in Eqs.~(\ref{gammadriver}) by 
$\partial/\partial t - \beta^a\partial_a$.\cite{vanMeter:2006vi,Bruegmann:2006at} 

The moving puncture method was discovered through careful numerical 
experimentation \cite{Campanelli:2005dd,Baker:2005vv} and has been applied successfully by a number of 
groups. (See, for example, 
Refs.~\cite{Herrmann:2006ks,Sperhake:2006cy,Bruegmann:2006at,Gonzalez:2006md,Campanelli:2006fy,Baker:2006kr,Thornburg:2007hu,Tichy:2007hk}.)
Recent analytical and numerical work on single, nonspinning (Schwarzschild) black holes was carried out with the 
purpose of clarifying  how the geometry evolves in the neighborhood of a 
puncture.\cite{Hannam:2006vv,Hannam:2006xw} These works focus on the behavior of the numerical slices in a 
three--dimensional evolution code, and find that the puncture evolves from a $\psi\sim 1/r$ 
singularity to a $1/\sqrt{r}$ singularity as the spacelike slice evolves to a stationary state. 
Because the physical metric behaves like $\psi^4(dr^2 + r^2 d\Omega)$ with $\psi\sim 1/\sqrt{r}$ 
near the puncture point $r=0$, the geometry represents 
a three--dimensional ``cylinder" of infinite proper length and finite areal radius. Stationary foliations with these 
properties have been described numerically and analytically.\cite{Estabrook:1973ue,Hannam:2006vv,Baumgarte:2007ht}

The purpose of this report is to extend these insights. I show that the time required for the puncture to evolve 
to a $1/\sqrt{r}$ singularity increases with increasing resolution. In the limit of infinite resolution the puncture would retain its $1/r$ character throughout any finite--time evolution. The behavior 
of the puncture that is seen in numerical studies can be misleading because the puncture region is 
not resolved. 

The geometrical picture of puncture evolution that is presented here applies to a single, 
nonrotating black hole. It is not clear whether the same description applies to 
black holes with spin and linear momentum. For puncture initial data 
the conformal factor diverges like $1/r$ at each puncture.\cite{Brandt:1997tf} This suggests that 
initially, the evolution of the geometry near each puncture point will be independent of the
spin and momentum. As the conformal factor changes, the geometry in the vicinity of a puncture could develop a dependence on spin and momentum. 

The discussion will be presented using the notation and terminology 
of the Kruskal--Szekeres diagram, shown in Fig.~1.\cite{Misner:1974qy}  
The Kruskal--Szekeres diagram represents the maximal analytic 
extension of a Schwarzschild black hole. 
\begin{figure}
\includegraphics[scale=0.8, viewport=200 300 300 500]{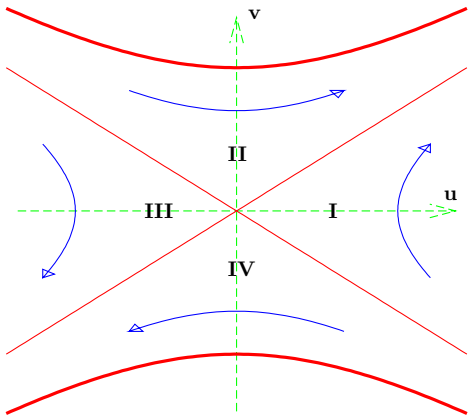}
\caption{Kruskal-Szekeres diagram of a Schwarzschild black hole.}
\end{figure}

The black hole 
interior is region II, and we live in the asymptotically flat exterior region I. Region III is the ``other'' asymptotically flat 
region, separated from I by a wormhole. The heavy curves are the future and past singularities. The axes 
$u$ and $v$ are Kruskal--Szekeres coordinates in which the metric reads 
\begin{equation}\label{Kruskalmetric}
  ds^2 = \frac{32 M^3}{R} e^{-R/2M} \left( -dv^2 + du^2 \right) + R^2 \, d\Omega^2 \ .
\end{equation}
Each point in the $u$--$v$ plane is a sphere of areal radius $R$, where $R$ is defined by 
\begin{equation}\label{arealradius}
  u^2 - v^2 = \left( \frac{R}{2M} - 1 \right) e^{R/2M} \ .
\end{equation}
The lines at $\sim 45^\circ$ are the horizons with $R = 2M$. The areal radius is greater than $2M$ in the exterior 
regions I and III. The areal radius is less than $2M$ in regions II and IV. 

The curved arrows in Fig.~1 show the directions of a Killing vector field, which is taken 
as future pointing in our region I. Note that any slice can be extended to a stationary foliation by Lie 
transport along the Killing vector field. In the Hamiltonian or 3+1 language, any spacelike slice can be evolved into 
a time--independent history if the lapse and shift are chosen such that the time flow 
vector field $t^\mu \equiv \alpha n^\mu + \beta^\mu$ coincides with the 
Killing vector field. (Here, $n^\mu$ is the unit timelike normal to the slices and $\beta^\mu$ is the shift 
vector expressed as a spacetime vector field.)  Note that  we cannot have 
a stationary foliation with $\alpha>0$ everywhere if the slices extend into both regions I and III. The foliations 
found in Refs.~\cite{Estabrook:1973ue,Hannam:2006vv,Baumgarte:2007ht} are examples of stationary foliations that begin 
in region II and end in region I, extending from 
$u\sim -\infty$, $v\sim\infty$ to $u\sim \infty$, $v\sim{\rm finite}$. For any such foliation whose 
slices asymptotically approach an $R = {\rm const}$  curve  in region II
(a hyperbola in Kruskal--Szekeres coordinates), the slice geometries will 
approach that of an infinitely long three--dimensional cylinder. 
The stationary foliation found in Refs.~\cite{Estabrook:1973ue,Baumgarte:2007ht} satisfies the maximality condition 
$K=0$, while the stationary foliation discussed in Ref.~\cite{Hannam:2006vv} satisfies the 
stationary ``1+log'' slicing condition $2\alpha K = \beta^a\partial_a \alpha$. 

The initial data for puncture evolution of a Schwarzschild black hole consist of the spatial metric
in isotropic coordinates, 
\begin{equation}\label{isotropicmetric}
  ds^2 = ( 1 + M/(2r))^4 \left( dr^2 + r^2 \, d\Omega^2 \right) \ ,
\end{equation}
and vanishing extrinsic curvature. These are the data for the $v=0$ slice of the Kruskal--Szekeres diagram. For the 
initial data the relationship between  $u$ and the isotropic radial coordinate $r$ is given by 
\begin{equation}\label{utor}
 u = \frac{1}{2} \left( \sqrt{\frac{2r}{M}} - \sqrt{\frac{M}{2r}} \right) \exp\left[ \frac{1}{8} \left( 
    \sqrt{\frac{2r}{M}} + \sqrt{\frac{M}{2r}} \right)^2 \right]  \ .
\end{equation}
The positive $u$ axis is covered by $M/2 \leq r < \infty$ and the negative $u$ axis is covered by 
$0 < r \leq M/2$. Initially 
the puncture point $r=0$ coincides with spacelike infinity in region III, $u = -\infty$. 

The key question is this. Can the initial data $v=0$ evolve to a slice that begins in region II and ends in 
region I? The answer is no. We can see this most clearly by keeping 
in mind that, in Hamiltonian language, the momentum constraint is the generator of spatial diffeomorphisms. As a 
consequence the time derivatives  in the 3+1 evolution equations appear only in the combination 
$\partial_\perp \equiv \partial/\partial t - {\cal L}_\beta$, where ${\cal L}_\beta$ is the Lie derivative 
along the shift vector $\beta^a$. Let us assume that 
the slicing condition has this same property: time derivatives appear only in the combination $\partial_\perp$. 
When this is the case, the evolution equations and slicing condition together determine the history of the (intrinsic and 
extrinsic) geometry of space. The  
lapse function specifies the point wise proper time separation between instances of space (or ``slices'') 
at successive times. The choice of shift vector does not affect the spatial geometry. The role of the shift vector 
is to determine how the spatial coordinates are carried from one slice to the next. 
In a finite difference calculation the shift vector determines how the distribution of grid points is 
carried from one slice to the next. 

Consider the evolution of the initial geometry represented by the $v=0$ slice using 1+log slicing, 
Eq.~(\ref{onepluslog}). We 
can ignore how the grid points are distributed initially, and how they are redistributed in time by the 
shift vector. Now observe that the maximally extended Schwarzschild black hole is symmetric under reflections about 
the Kruskal--Szekeres $v$ axis. Thus, the initial metric expressed in Kruskal--Szekeres coordinates (the metric 
of Eq.~(\ref{Kruskalmetric}) with $v = 0$) is invariant 
under $u \to -u$. The initial metric expressed in isotropic coordinates Eq.~(\ref{isotropicmetric}) is 
invariant under $r \to M^2/(4r)$.
This reflection symmetry implies that the initial spatial geometry in region III 
is identical to the initial spatial geometry in region I. As long as the initial choice of lapse function is also 
reflection symmetric, then the geometry will remain symmetric as it evolves in time. 
If the part of the slice that extends to spatial infinity in region I does not evolve into region II, then the 
part of the slice that extends to spatial infinity in region III also does not evolve into region II. 

In fact, neither end of the initial data slice $v=0$ can evolve entirely into the black hole interior (region II) 
in finite time. To see why this is the case, consider a sequence of spacelike slices whose members are separated 
by finite proper time, and whose 
initial slice is $v=0$. Starting from a fixed point in region III on the initial slice, we can define a timelike 
trajectory of finite proper length that is orthogonal to each of the slices. 
Now take the limit as the fixed initial point moves to spacelike 
infinity. A timelike trajectory cannot extend from spacelike infinity to the black hole interior in finite proper 
time. This is 
clear from the Penrose diagram \cite{Misner:1974qy}, which shows that a 
curve connecting spacelike infinity to the black hole interior 
must be null or somewhere spacelike. It follows 
that  the timelike trajectory remains in region III. The spacelike slices must also extend into region III. 

The argument above shows that the puncture point, 
which begins at spacelike infinity in region III, must remain in region III throughout its evolution. All of the 
slices evolved from $v=0$ begin in region III and end in region I.  With a positive lapse these slices cannot 
reach a stationary state. 

These results are confirmed by numerical tests. The tests described below 
are carried out with a BSSN code that assumes spherical 
symmetry. The code is based on a generalization of the BSSN formulation that does not assume 
$\det({\tilde g}_{ab})=1$.\cite{Brown:2005aq} The evolved data include the components ${\tilde g}_{rr}$ 
and ${\tilde g}_{\theta\theta}$ of the conformal metric, 
the ``chi version'' of the conformal factor $\chi \equiv 1/\psi^4$, the trace of the extrinsic curvature $K$, 
the component ${\tilde A}_{rr}$ of the trace--free part of the extrinsic curvature, and 
the conformal connection function ${\tilde\Gamma}^r$. The lapse function $\alpha$ is determined 
by the 1+log slicing condition Eq.~(\ref{onepluslog}). Each of these variables, along with the shift vector 
component $\beta^r$, is  a function of time $t$ and radial coordinate $r$. The initial metric is given in 
Eq.~(\ref{isotropicmetric}). The finite difference grid is cell centered 
with grid points $r_j = (j - 1/2)\Delta r$ for $j=1,2,\ldots$. The puncture is located at $r=0$.
The code also tracks the Kruskal--Szekeres coordinates $u$ and $v$ of each grid point; this allows the slices 
to be displayed in a  Kruskal--Szekeres or Penrose diagram. 
Further details of the 1-D code are described in Ref.~\cite{Brown:BSSNss}.


Figure 2 shows the slice at time $t=3M$ as a curve just below the future singularity on the Kruskal--Szekeres diagram. 
The slice begins in region III and ends in region I. This slice was obtained by evolving the initial data $v=0$ with 
initial lapse function $\alpha = 1$. For this simulation the shift vector was set to zero for all time. 
The reflection symmetry about the $v$--axis 
is evident. The heavy dots in Fig.~2 show the locations of grid points obtained from a second simulation, 
identical to the first except for the use of the $\Gamma$--driver shift condition. 
For clarity of presentation only every other grid point 
is displayed, beginning with grid point $j=2$. 
The resolution used for both of these runs was $\Delta r = M/50$; this is comparable to the highest 
resolutions used in current 3D codes. 
\begin{figure}
\includegraphics[scale=0.85, viewport=180 300 380 480]{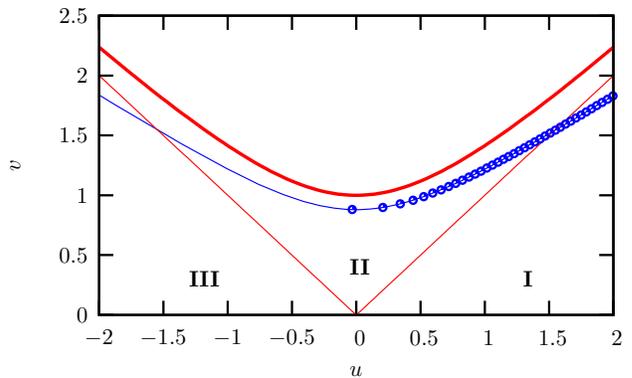}
\caption{Slice at $t = 3M$ for vanishing shift (solid curve) and 
$\Gamma$--driver shift (dots).}  
\end{figure}

With the $\Gamma$--driver shift, the left--hand side of the Kruskal--Szekeres diagram is almost completely devoid of 
grid points by $t=3M$. The asymmetry in the distribution of grid points occurs for two reasons. First, the 
grid points in the initial data are distributed uniformly 
in $r$, not  in $u$. With an outer boundary at, say, $100M$, only $0.5\%$ of the grid points 
start in the left half of the Kruskal--Szekeres diagram. 
Second, with the $\Gamma$--driver condition the shift vector quickly develops $\beta^r \sim r$ behavior near the puncture. 
Since the conformal factor diverges like $1/r$, the magnitude of the shift vector behaves like 
$\sqrt{\beta^r \psi^4 {\tilde g}_{rr} \beta^r} \sim 1/r$. Due to its large magnitude at small $r$, the shift vector quickly drives the 
grid points in region III toward the black hole interior. 

The spherically symmetric code differs from current three--dimensional black hole codes in one 
important respect: with the spherically symmetric code the 
puncture point $r=0$ is a boundary of the computational domain. In most contexts, 
the choice of boundary conditions will affect the evolution of the fields in the bulk. For puncture 
evolution, however, the results are found to be 
insensitive to the choice of boundary conditions at $r=0$. 
This is most likely due to that fact that, with the $\Gamma$--driver shift condition, the 
grid points near the puncture quickly acquire superluminal speeds as they are drawn into 
the black hole interior. The grid points continue to evolve along spacelike 
trajectories that approximate the orbits of the Killing vector field. It follows that at the 
boundary $r=0$ all of the physical characteristics are directed from interior to exterior 
(toward the left in the Kruskal--Szekeres diagram). As a result one would not expect conditions
imposed at $r=0$ to influence the evolution of the geometry. 

With the spherically symmetric code we can easily increase 
the resolution well beyond the value used in the simulations of Fig.~2. This is an effective way to
counteract the sparsity of grid points in region III in the initial data. For example, with  $\Delta r = M/50$ and the outer boundary at $100M$ there are only 
$25$ grid points with initial values $u < 0$. We can increase this by a factor of ten or more and still maintain 
reasonably short run times. However, a straightforward increase in resolution is not very effective in counteracting the 
movement  of grid points caused by the $\Gamma$--driver shift condition. At resolutions as high as $\Delta r = M/6400$
the shift vector still drives all of the grid points into the black hole interior within a time of a few $M$. See Fig.~3. After all of the 
grid points have left region III, the lapse and shift can settle into a stationary state where the time 
flow vector field $t^\mu = \alpha n^\mu + \beta^\mu$ coincides with the Killing vector field. 

For higher resolutions the time required for all of the grid points from region III to be driven into 
the black hole interior increases, as seen in Fig.~3. 
\begin{figure}
\includegraphics[scale=0.85, viewport=180 300 380 480]{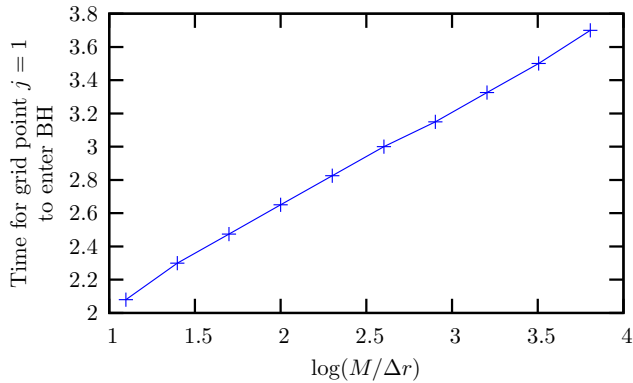}
\caption{The vertical scale is the time required (in units of $M$) for all of the grid points that begin in region III 
to evolve into the black hole interior, region II, with the $\Gamma$--driver shift condition. The horizontal scale is the 
common logarithm of $M/\Delta r$. The resolutions used for this 
graph differ by powers of $2$, ranging from  $\Delta r = M/12.5$ to $\Delta r = M/6400$.}
\end{figure}
In the limit of infinite resolution, the grid points at any finite time would stretch from region III to region I,  
crossing the black hole interior like the curve in Fig.~2.  At infinite resolution the slices never reach a stationary state. 
However, the time $T$ required for the region III grid points to evolve into region II increases very slowly with resolution, 
roughly like $T/M \sim 0.58\log(M/\Delta r) + 1.5$. We can compare this to the time scale for the lapse and shift to settle 
to a stationary state---at resolutions typical of current codes this is about $50M$. According to the heuristic expression for 
$T/M$, an extraordinarily high resolution of 
$\Delta r \sim M/10^{83}$ would be required for grid points to remain in region III beyond $t\sim 50M$.

One of the important properties of the 1+log slicing condition is that the lapse function is driven to zero in the black hole 
interior, so  the slices  naturally avoid the physical 
singularity. Figure 4 shows the common logarithm of the lapse at time $t=30M$ 
plotted as a function of areal radius for both $u>0$ (right half of the graph) and $u<0$ (left half of the graph). 
\begin{figure}
\includegraphics[scale=0.85, viewport=180 300 380 480]{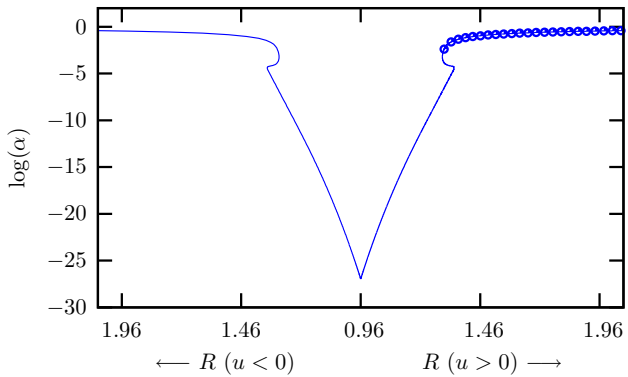}
\caption{Logarithm of the lapse $\alpha$ as a function of areal radius $R$ (in units of $M$) at time $t=30M$ for vanishing 
shift (solid curve) and $\Gamma$--driver shift (dots).}
\end{figure}
The 
continuous curve was obtained from a simulation with vanishing shift vector and resolution $\Delta r = M/800$. The dots 
were obtained from a simulation with the $\Gamma$--driver shift condition and resolution 
$\Delta r = M/50$. (For clarity, only 
every other grid point is displayed.) Again we see that the evolution of the complete slice is symmetric under reflections 
$u \to -u$, but for the simulation with $\Gamma$--driver shift the distribution of grid points is highly asymmetric. 

With the moving puncture method the region inside the black hole horizon is not completely resolved. 
As a matter of principle, 
the puncture point remains at spacelike infinity in region III throughout its evolution. However, 
with the $\Gamma$--driver condition the shift vector quickly drives all of the region III grid 
points into the black hole interior. The resulting slices are incomplete; the grid points terminate in region II due to 
lack of resolution.  The lapse function and shift vector adjust to bring the data  
to a stationary state. The stationary state is a portion of a slice that extends from $u\sim -\infty$, $v\sim\infty$ in 
region II to $u\sim\infty$, $v\sim{\rm finite}$ in region I. 

The lack of resolution that characterizes the puncture method is not likely to cause problems for finite difference codes at 
any reasonable resolution. In fact, the results here suggest that 
puncture evolution should be viewed as a type of  ``natural excision". With black hole excision one 
intentionally removes from the computational domain a set of grid points inside the horizon (see, for 
example, Ref.~\cite{Alcubierre:2004bm}). In the case of 
puncture evolution the removal of grid points occurs naturally. 
The advantage over the traditional excision method is that puncture evolution 
does not require the introduction of an interior boundary where boundary conditions and special finite difference 
stencils must be applied. 

I would like to thank Mark Hannam, Pablo Laguna, Niall O'Murchadha and Bernard Whiting for helpful discussions. 
This work was supported by NSF grant PHY--0600402.

\bibliography{references}

\end{document}